**Bioinspired Microactuators Fabricated via One-step Meniscus-guided 3D Nanoprinting**

*Seong-Jae Eom, Vasanthan Devaraj*, Sunghyun Kwak, Hyeon-Seok Seo, Thomas Zentgraf*, Won-Geun Kim*, Jong-Min Lee**

Seong-Jae Eom, Jong-Min Lee
School of Semiconductor Display Technology, Hallym University, Chuncheon 24252, Republic of Korea
E-mail: jmlee@hallym.ac.kr

Vasanthan Devaraj, Thomas Zentgraf
Department of Physics, Paderborn University, 33098 Paderborn, Germany
Institute for Photonic Quantum Systems (PhoQS), Paderborn University, 33098 Paderborn, Germany
E-mail: vasanthan.devaraj@uni-paderborn.de, thomas.zentgraf@uni-paderborn.de

Vasanthan Devaraj
Center for Smart Engineering Materials, New York University Abu Dhabi, 129188 Abu Dhabi, United Arab Emirates
Email: vd2629@nyu.edu

Sunghyun Kwak, Won-Geun Kim
Department of Optical Engineering, Kumoh National Institute of Technology, Gumi 39177, Republic of Korea
E-mail: wgkim@kumoh.ac.kr

Hyeon-Seok Seo, Jong-Min Lee
School of Nano Convergence Technology & Nano Convergence Technology Center, Hallym University, Chuncheon 24252, Republic of Korea

Won-Geun Kim
Department of Physics, Kumoh National Institute of Technology, Gumi 39177, Republic of Korea

**Abstract text.**

We introduce a bioinspired microactuator fabricated via a one-step meniscus-guided 3D nanoprinting technique. This technique directly produces a freestanding three-dimensional composite architecture by integrating a rigid nanoparticle framework with a hygroscopic polymer matrix in a single, assembly-free process. Inspired by the graded hard–soft interface of an insect exoskeletal joint, our design synergistically combines load-bearing strength and humidity-driven swelling in a single microscale pillar. Comprehensive experiments, including comparative control structures and microscopic analysis, elucidated the actuation mechanism: the nanoparticle scaffold provides mechanical support while the polymeric phase provides volumetric expansion, yielding large reversible elongation under humidity with preserved structural integrity. The resulting composite pillar exhibits muscle-like performance at the microscale, lifting loads orders of magnitude heavier than its own weight and sustaining extensive humidity cycling with negligible performance degradation. Variant geometries such as hinged pillars demonstrate how axial swelling can be converted into bending motion, and an optical configuration shows how actuation can modulate reflected light signals. This versatile nanoprinting approach offers a simple route to architect bioinspired soft microactuators with high strength, durability, and multifunctional responsiveness, paving the way for integration into future micro-robotic, sensing, and hybrid electronic systems.

## 1. Introduction

Micro-scale actuators are fundamental components for microsystems and microrobots, enabling motion and force generation at tiny scales.[1-3] However, creating effective actuators at the microscale remains challenging, as conventional actuation mechanisms often require complex, multi-step fabrication processes and the integration of multiple materials.[4-9] These constraints significantly limit structural complexity, scalability, and mechanical robustness at small length scales.

In contrast, natural systems demonstrate remarkable mechanical performance despite their small size.[10,11] A representative example is the neck joint of an ant, which consists of rigid chitinous exoskeletal plates interconnected by compliant soft tissue, yet can withstand forces several thousand times the ant's body weight.[12,13] Biomechanical studies have revealed that this extraordinary load-bearing capability does not arise from muscle strength alone, but from a graded hard–soft interface between the exoskeleton and the arthrodial membrane.[14-16] Rather than forming an abrupt material junction, ant neck joints exhibit a finely interdigitated transition that effectively redistributes stress and suppresses localized failure, thereby enhancing mechanical robustness despite the relatively modest stiffness of the soft tissue.[12] This bioinspired design principle suggests that integrating rigid and deformable components through a gradual material transition could enable microactuators with exceptional strength-to-weight ratios and durability.[17,18]

Among various actuation strategies, humidity-responsive polymers have attracted attention due to their ability to swell and shrink in response to environmental moisture without requiring electrical input.[19-22] Humidity-driven actuators have been extensively studied in bilayer configurations, where differential expansion induces bending motion.[23-25] Although such bilayer actuators can exhibit large deformations, they often suffer from weak interfacial adhesion, delamination under cyclic operation, and limited deformation modes restricted to simple curvature.[26] To address these issues, monolithic actuators based on single materials or single layers have been explored. For example, Pan et al. demonstrated a humidity-responsive actuator using a single graphene oxide film with asymmetric internal structure, achieving bending angles up to 1800°.[27] Troyano et al. reported polymer films embedded with metal–organic framework (MOF) crystals that exhibited reversible shape-memory deformation under humidity variation.[28,29] Despite these advances, existing humidity-responsive actuators remain largely planar, and a strategy to fabricate truly three-dimensional, freestanding microactuator architectures in a single step is still lacking.[30] Additive manufacturing provides a promising route to overcome these limitations by enabling the direct fabrication of complex three-dimensional microstructures. In particular, meniscus-guided 3D printing has recently emerged as an accessible microscale printing technique capable of writing freestanding micro- and nanostructures directly from solution in a single step.[31–36] In this approach, a nanoparticle ink confined within a liquid meniscus is continuously deposited through coordinated solvent evaporation and meniscus translation, enabling precise vertical growth and layer-by-layer construction. The method offers sub-micron resolution, high aspect ratios, and broad material compatibility, making it well suited for constructing bioinspired composite architectures without the need for multi-step processing.[37]

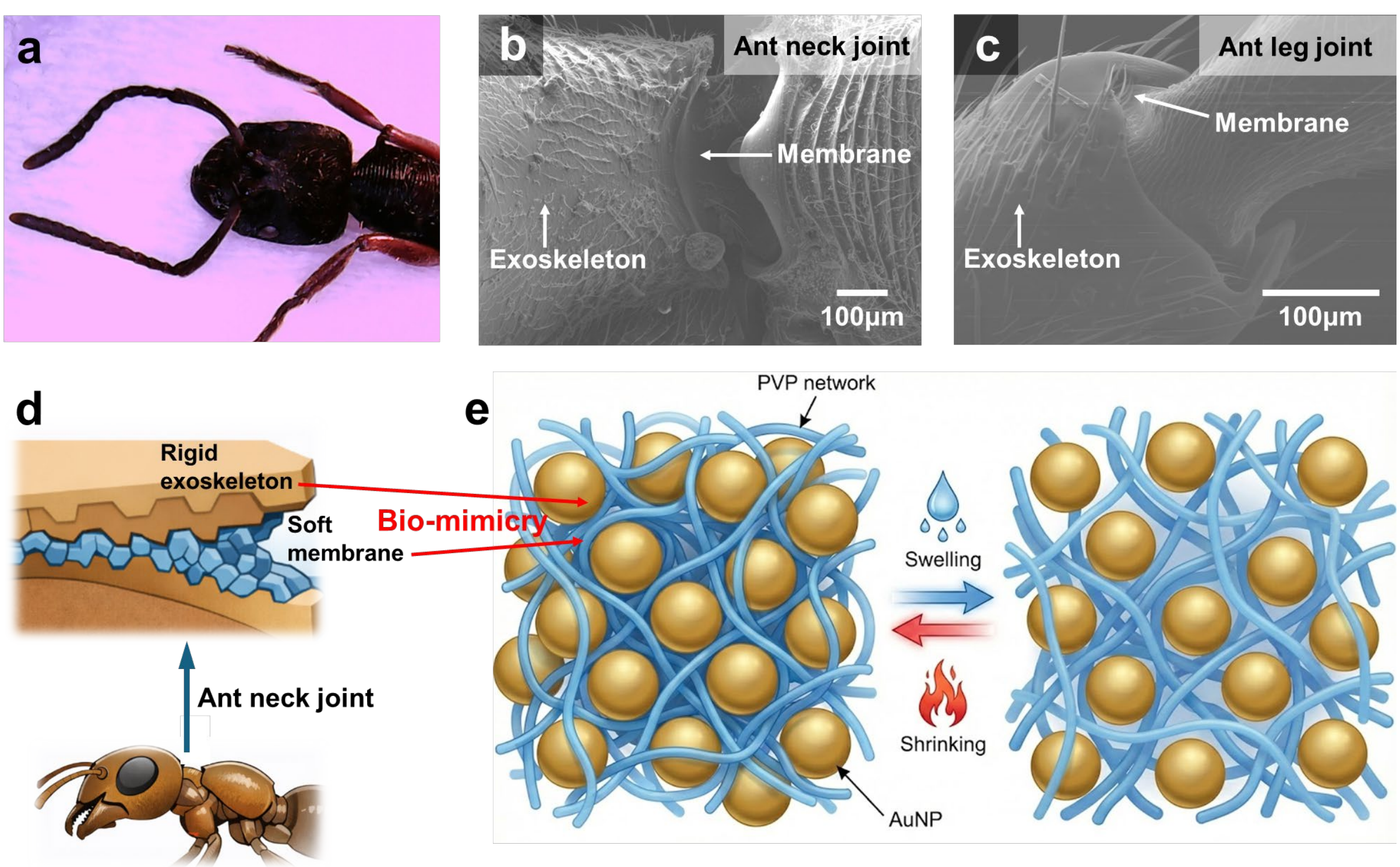

**Figure 1.** (a) Optical photograph of an ant highlighting its articulated body structure. (b) Scanning electron microscope (SEM) image of an ant neck joint, showing a rigid exoskeleton connected by a compliant membrane. (c) SEM image of an ant leg joint, illustrating a similar hard–soft joint architecture composed of a stiff exoskeleton and a flexible membrane. (d) Schematic illustration of a biological hard–soft interface, consisting of a rigid exoskeleton layer and an underlying soft membrane. (e) Conceptual schematic of the bioinspired microactuator, in which gold nanoparticles are embedded within a polymer network to form a composite structure responsive to external stimuli such as humidity and heat.

Here, we report a bioinspired humidity-responsive microactuator fabricated via one-step meniscus-guided 3D printing. Our design is inspired by the graded, hard–soft interface of ant neck joints. Optical microscopy reveals the compact morphology of the ant head–thorax junction that supports substantial mechanical loads (Fig. 1a), while scanning electron microscopy resolves rigid exoskeletal plates interconnected by a compliant arthrodial membrane with complex interfacial geometry (Fig. 1b and 1c). Translating this biomechanical principle, we designed an artificial microscale actuator in which a load-bearing gold nanoparticle (AuNP) network replaces the rigid exoskeleton, and a hygroscopic polyvinylpyrrolidone (PVP) phase serves as a deformable component analogous to the ant's arthrodial membrane (Fig. 1d). The AuNPs and PVP form an interpenetrating composite structure, establishing a gradual mechanical transition rather than an abrupt interface. This bioinspired material gradient enables efficient stress redistribution during actuation, allowing large reversible deformation while preserving structural integrity under load.

Importantly, PVP is widely used as a stabilizing ligand for gold nanoparticles in colloidal synthesis. By exploiting the intrinsic hygroscopic nature of PVP already present on the nanoparticle surface, our approach eliminates the need to formulate additional hydrogel–nanoparticle composite inks. This ligand-enabled actuation strategy significantly simplifies material preparation and enables direct utilization of well-established AuNP inks, providing a versatile platform for one-step fabrication of bioinspired microactuators. The resulting microactuators are freestanding, truly three-dimensional, and ready for use immediately after printing, without any post-assembly or transfer steps. We demonstrate that these micropillars sustain repeatable humidity-driven actuation without performance degradation, while lifting loads hundreds of times greater than their own weight. By combining bioinspired graded composite design with scalable one-step 3D nanoprinting, this work establishes a new paradigm for microscale actuator fabrication and opens opportunities for applications in microrobotics, sensors, and adaptive microdevices.

## 2. Results and Discussion

To fabricate the bioinspired microactuator, we employed meniscus-guided 3D printing, a direct-write method that deposits solutes by leveraging rapid solvent evaporation from a femtoliter (fL)-scale meniscus formed upon contact between a glass micropipette (loaded with ink solution) and a substrate. This approach enables high-resolution printing by scaling down from micron to sub-micron scale (hundreds-of-nanometers) regime. Figure 2a presents a schematic illustration of the AuNP ink used in this study, together with an optical microscope image of the AuNP solution loaded at the pipette tip. The AuNPs have a diameter of 10 nm and are coated with PVP, which serves both as a dispersant and as a hygroscopic component that contributes to actuation. The glass micropipette used as the printing nozzle was prepared using a commercial micropipette puller, producing a nozzle diameter of approximately 10 μm. When the opening on the opposite side of the micropipette is immersed in the AuNP solution, the ink is transported to the nozzle tip by capillary action.

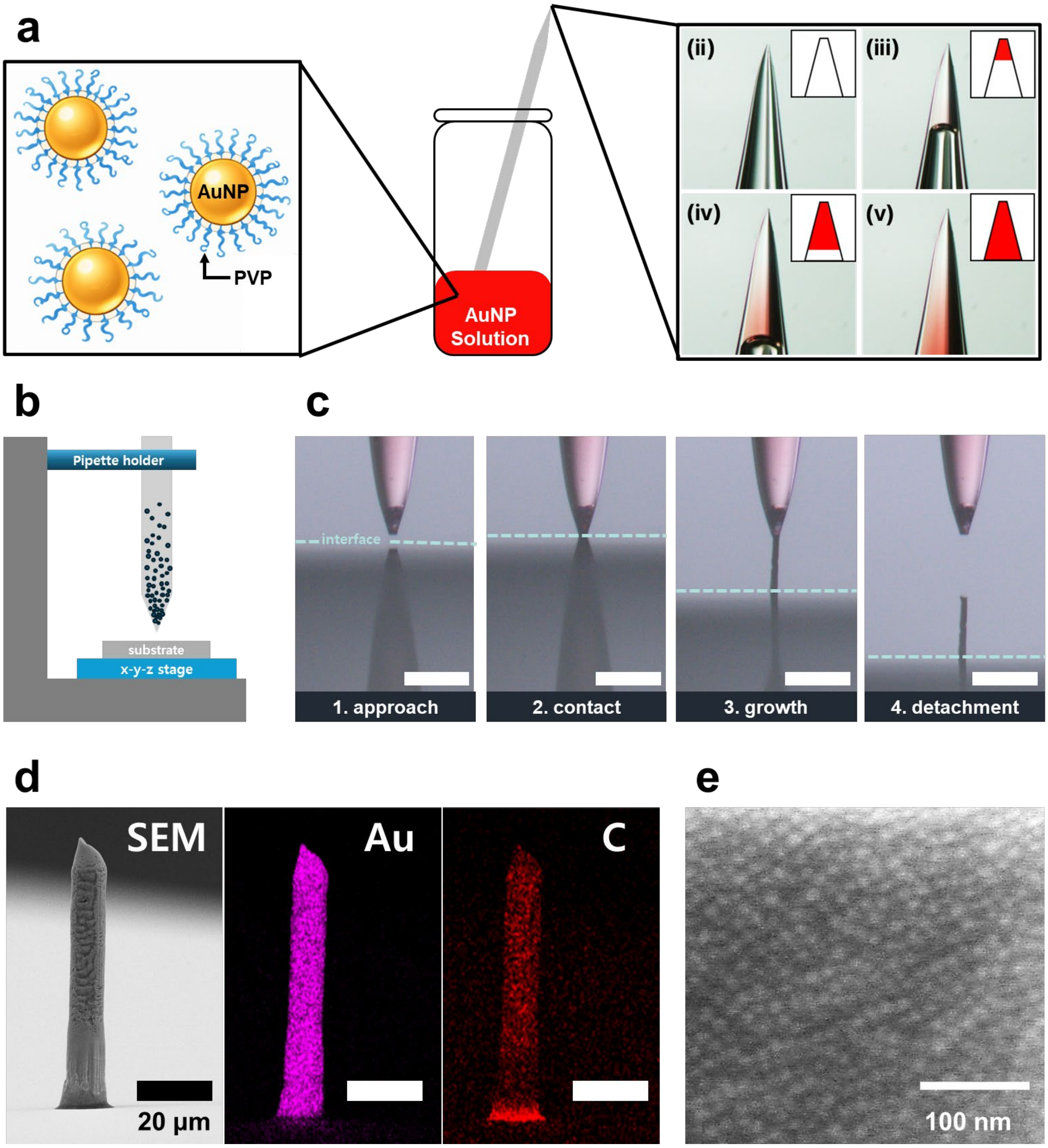

**Figure 2.** (a) Schematic illustration of PVP-coated AuNPs used as the printing ink, together with optical microscope images of a glass micropipette filled with the AuNP solution and representative micropipette tips with different nozzle geometries prepared using a micropipette

puller. (b) Schematic illustration of the meniscus-guided 3D nanoprinting setup. (c) Optical microscope images showing the meniscus-guided printing process. (d) SEM image of a representative freestanding microactuator pillar and the corresponding EDS elemental maps of Au and C. (e) High-magnification SEM image revealing the nanoscale composite microstructure composed of densely packed AuNPs embedded in a polymer matrix.

Figure 2b illustrates the experimental setup for meniscus-guided 3D printing. The ink-loaded micropipette is mounted on a custom-built pipette holder, while the substrate is placed on a motorized x–y–z stage. The approach/contact of the nozzle to the substrate and the subsequent printing (growth) of the microstructures are executed by controlling the stage. In addition, a simple optical microscope module is integrated into the setup to monitor nozzle–substrate contact and real-time structure growth. Figure 2c shows optical microscope snapshots of the full printing sequence. The substrate is approached toward and brought into contact with the micropipette fixed on the holder by operating the motorized stage. Upon contact, an ultrasmall fL meniscus forms between the nozzle and the substrate due to the micrometer-sized nozzle aperture. Because of the extremely high surface-to-volume ratio of the fL meniscus, solvent evaporation proceeds rapidly, enabling efficient solidification of the solute. Immediately after contact, lowering the motorized stage at a controlled speed of 300 nm/sec allows the bioinspired microactuator to grow vertically. After reaching the desired height, the stage is moved downward rapidly (≈ 1 mm/sec) to detach the pipette from the printed structure. Figure 2d presents scanning electron microscopy (SEM) images and energy dispersive x-ray spectroscopy (EDS) results of the printed Au–PVP bioinspired microactuator. The printed structure exhibits a freestanding pillar geometry with a diameter of ~10 μm and a length of ~70 μm. The EDS maps confirm that AuNPs and PVP (carbon signal) are distributed uniformly throughout the entire microactuator. Furthermore, field-emission SEM (FE-SEM) imaging (Fig. 2e) reveals that a continuous PVP network is formed across the structure, with AuNPs uniformly embedded within the network, consistent with the schematic illustration in Fig. 1d.

Owing to the hygroscopic nature of PVP, the bioinspired microactuator can undergo humidity-dependent shape changes. To elucidate the actuation mechanism and decouple the roles of (i) hygroscopic polymer swelling, (ii) a non-swellable nanoparticle framework, and (iii) a purely metallic skeleton, we prepared four types of freestanding pillars and exposed them to high humidity (RH = 98%): a PVP-only pillar printed from PVP solution (Fig. 3a), a $SiO_2$ nanoparticle ($SiO_2$NP) pillar printed from silanol-terminated (non-swellable material) silica nanoparticles (Fig. 3b), an Au sintered pillar obtained by printing PVP-coated AuNPs followed by thermal treatment to remove PVP and sinter the Au framework (Fig. 3c), and the bioinspired microactuator printed directly from PVP-coated AuNPs (Fig. 3d). Under RH = 98%, the PVP-only pillar did not exhibit repeatable actuation; instead, it underwent humidity-induced softening and ultimately collapsed/spread, losing its freestanding geometry. In contrast, neither the $SiO_2$ NP pillar nor the Au sintered pillar exhibited a measurable actuation response under the same conditions, indicating that rigid inorganic/metallic frameworks lacking a hygroscopic phase remain largely shape-invariant, whereas a hygroscopic polymer phase alone – without structural reinforcement – tends to fail by collapse. Notably, the PVP-coated AuNP microactuator displayed a pronounced humidity-triggered actuation mode relative to ambient conditions (Supplementary Movie S1), demonstrating that robust freestanding actuation arises from the synergy between PVP-driven moisture uptake and a nanoparticle-based load-bearing architecture.

Mechanistically, at high humidity PVP readily absorbs water and becomes strongly plasticized, which reduces the effective glass transition temperature and mechanical stiffness while increasing chain mobility. For PVP-only pillars, this promotes viscoelastic creep/flow; once the softened polymer yields, deformation preferentially proceeds via wetting-assisted flattening (i.e., increasing substrate contact area), resulting in collapse and spreading rather than shape-preserving deformation. By contrast, in AuNP–PVP composite pillars, rapid solvent evaporation during meniscus-guided printing concentrates the AuNPs, enabling the formation of a percolated, load-bearing nanoparticle skeleton that efficiently transfers stress and suppresses large-scale viscous deformation. In addition, adsorption of PVP on Au surfaces creates a constrained polymer fraction that further inhibits humidity-induced flow. Consequently, although moisture uptake still induces swelling of the PVP phase, the deformation is expressed as constrained, shape-preserving swelling rather than structural failure. Here, a key boundary condition is that the pillar base is anchored to the substrate after printing (approximately clamped or partially clamped, depending on adhesion and interfacial friction), which restricts lateral slip and radial relaxation near the base. As a result, volumetric swelling is redirected into anisotropic actuation, manifested as a large increase in pillar height accompanied by only a modest increase in diameter.

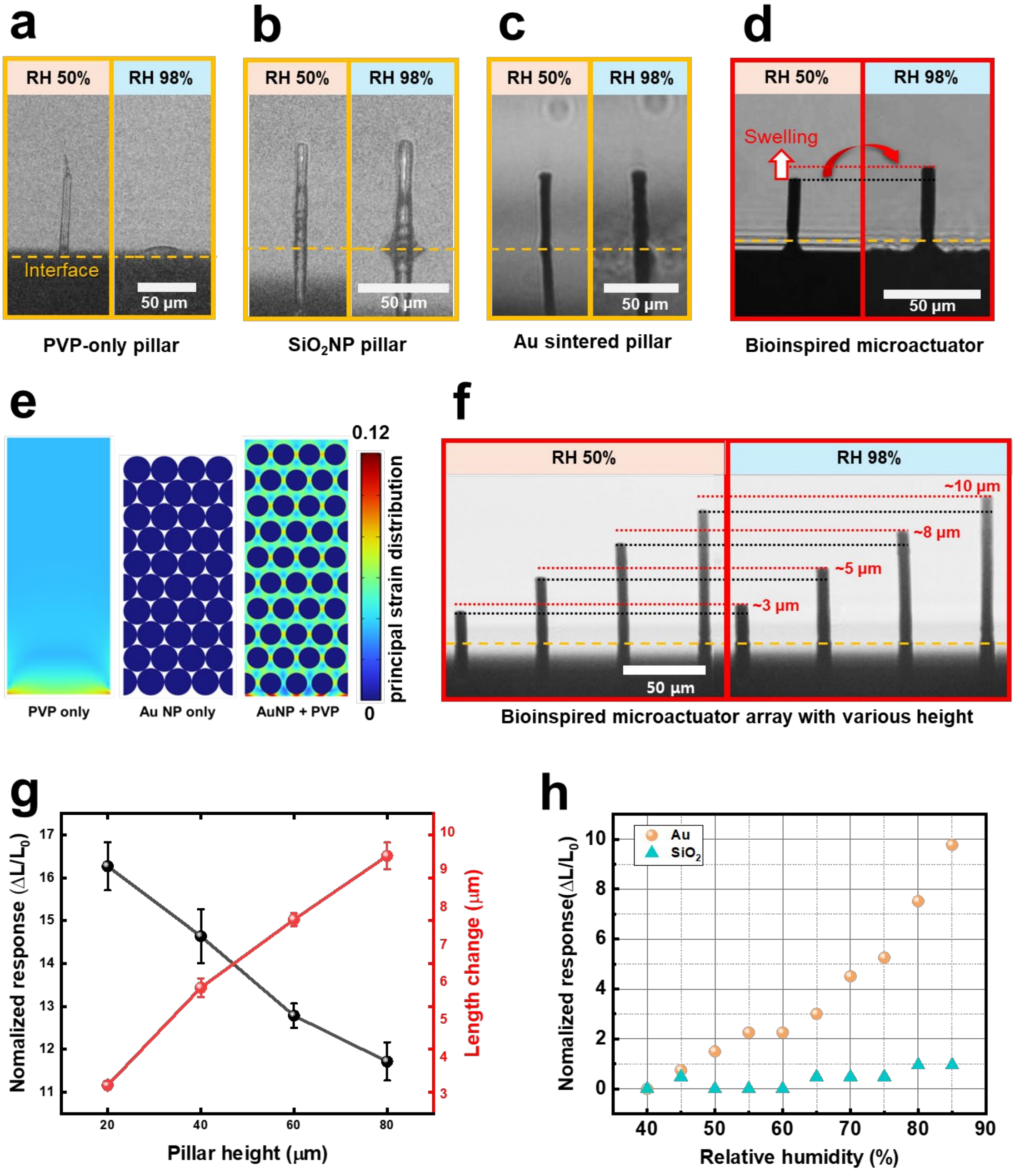

**Figure 3.** (a) Optical microscope images of a PVP-only pillar under low (RH 50%) and high (RH 98%) humidity conditions, showing humidity-induced softening and loss of the freestanding geometry. (b) Optical microscope images of a $SiO_2$ nanoparticle ($SiO_2$NP) pillar under low and high humidity, showing negligible dimensional change. (c) Optical microscope images of an Au sintered pillar under low and high humidity, indicating humidity-insensitive behavior. (d) Optical microscope images of a bioinspired microactuator under low and high humidity, showing reversible axial elongation upon humidity increase. (e) Principal strain distributions for PVP-only, AuNP-only, and AuNP–PVP composite pillars under high humidity (RH = 98%) with a clamped base, highlighting distinct deformation modes and constrained axial elongation in the bioinspired microactuator. The color (blue to red) strength (principal

strain distribution) is a dimensionless unit. (f) Optical microscope images of bioinspired microactuators with different pillar heights under low and high humidity, illustrating height-dependent elongation behavior. (g) Quantitative plot of absolute length change and normalized response ($\Delta L/L_0$) as a function of pillar height for bioinspired microactuators under high humidity. (h) Comparison of humidity-induced length change for AuNP–PVP microactuators and $SiO_2$NP pillars, highlighting the distinct responses of the two structures.

To quantitatively support the proposed actuation mechanism, finite-element simulations were performed to analyze humidity-induced deformation and strain distributions under clamped boundary conditions (Fig. 3e). Although the actuator possesses a three-dimensional pillar geometry, the deformation behavior is governed by axial symmetry and base clamping; therefore, a two-dimensional model sufficiently captures the essential physics of strain localization and boundary-condition-induced deformation. The simulations reveal that, in the PVP-only pillar, hygroscopic swelling leads to pronounced localization of the principal strain near the substrate–pillar interface due to the absence of mechanical reinforcement, providing a mechanical basis for the experimentally observed collapse under high humidity (Fig. 3a). In contrast, the AuNP-only pillar exhibits negligible deformation across the entire structure, as no hygroscopic phase is present to generate swelling-induced eigenstrain, consistent with its humidity-inactive behavior observed experimentally (Fig. 3c). By comparison, in the AuNP–PVP composite pillar (bioinspired microactuator), hygroscopic swelling is not confined to a single region but is spatially redistributed throughout the nanoparticle network, resulting in constrained and stable axial elongation. While the absolute displacement predicted by the static elastic simulations remains on the nanometer scale, this does not contradict the micrometer-scale actuation observed experimentally. Rather, the simulations are intended to elucidate the mode and spatial distribution of swelling-induced strain, rather than to reproduce the absolute magnitude of deformation. Collectively, these results demonstrate that functional actuation is governed not by the magnitude of hygroscopic swelling alone, but by how the swelling strain is constrained and guided by the nanoparticle framework and the clamped boundary condition at the pillar base.

High-humidity exposure (RH = 98%) induces pronounced axial elongation of the AuNP–PVP microactuators, and the absolute length change ($\Delta L$) increases with the as-printed pillar height (Fig. 3f). However, the normalized response ($\Delta L/L0$, length change rate) decreases for longer pillars, indicating that the actuation strain is not uniform along the pillar axis (Fig. 3g).

SEM/EDS analysis (Fig. 2d) reveals an AuNP-rich basal region, whereas the mid-to-upper sections exhibit a more uniform AuNP–PVP distribution, suggesting that the humidity-responsive deformation primarily originates from the polymer-rich, compositionally homogeneous region. Notably, a finite delay is typically required between ink loading into the glass micropipette and the initiation of printing; during this waiting period, solvent evaporation occurs at the nozzle tip, locally concentrating AuNPs and thereby forming an AuNP-rich region at the pillar base. The Au-rich base forms a mechanically stiff, percolated skeleton with reduced hygroscopic contribution, thereby acting as a low-strain anchor layer. In addition, because the pillar base is anchored to the substrate (approximately clamped or partially clamped), lateral slip and radial relaxation near the base are strongly restricted, and a portion of the swelling-induced driving force is consumed by internal constraint rather than converted into free axial extension. Although such basal constraints - arising from both the boundary condition and the high local AuNP density - can reduce the normalized response, they are expected to enhance the overall durability of the actuator by suppressing viscous flow, collapse, and structural instability under high humidity. As a result, increasing pillar height adds more actuating material and thus increases $\Delta L$, while the growing influence of the stiff, substrate-clamped basal constraint reduces the average axial strain, leading to a smaller $\Delta L/L0$ for longer pillars. Although a proportional increase in normalized actuation strain with pillar height may be inferred from the increased swelling volume, the presence of an AuNP-rich stiff basal layer and substrate clamping imposes a mechanical constraint that redistributes the swelling strain. As a result, the absolute elongation increases with height, whereas the normalized response decreases due to constraint-dominated deformation. Additionally, while the silanol-terminated $SiO_2$NPs - serving as a non-swellable material - exhibit little to no mechanical response to changes in relative humidity, the bioinspired microactuator shows an approximately linear response that scales with relative humidity (Fig. 3h).

Figure 4a shows that the actuator's humidity-driven elongation remains essentially unchanged over ~100 alternating cycles between RH 98% and RH 50%. In the main plot, the black trace of length change is reproduced cycle after cycle with no systematic drift, and the zoomed insets for cycles 0–10 and 90–100 overlay closely. This indicates negligible fatigue: the peak extension and baseline return in the 100th cycle is virtually identical to those in the first few cycles. Quantitatively, no appreciable reduction in actuation amplitude is observed. In other words, the pillar continues to swell and shrink by ~10–15% of its length each humidity cycle without loss of performance. Such robust cycling is consistent with the bioinspired design: the rigid AuNP network maintains structural integrity while the PVP can reversibly uptake and release water.

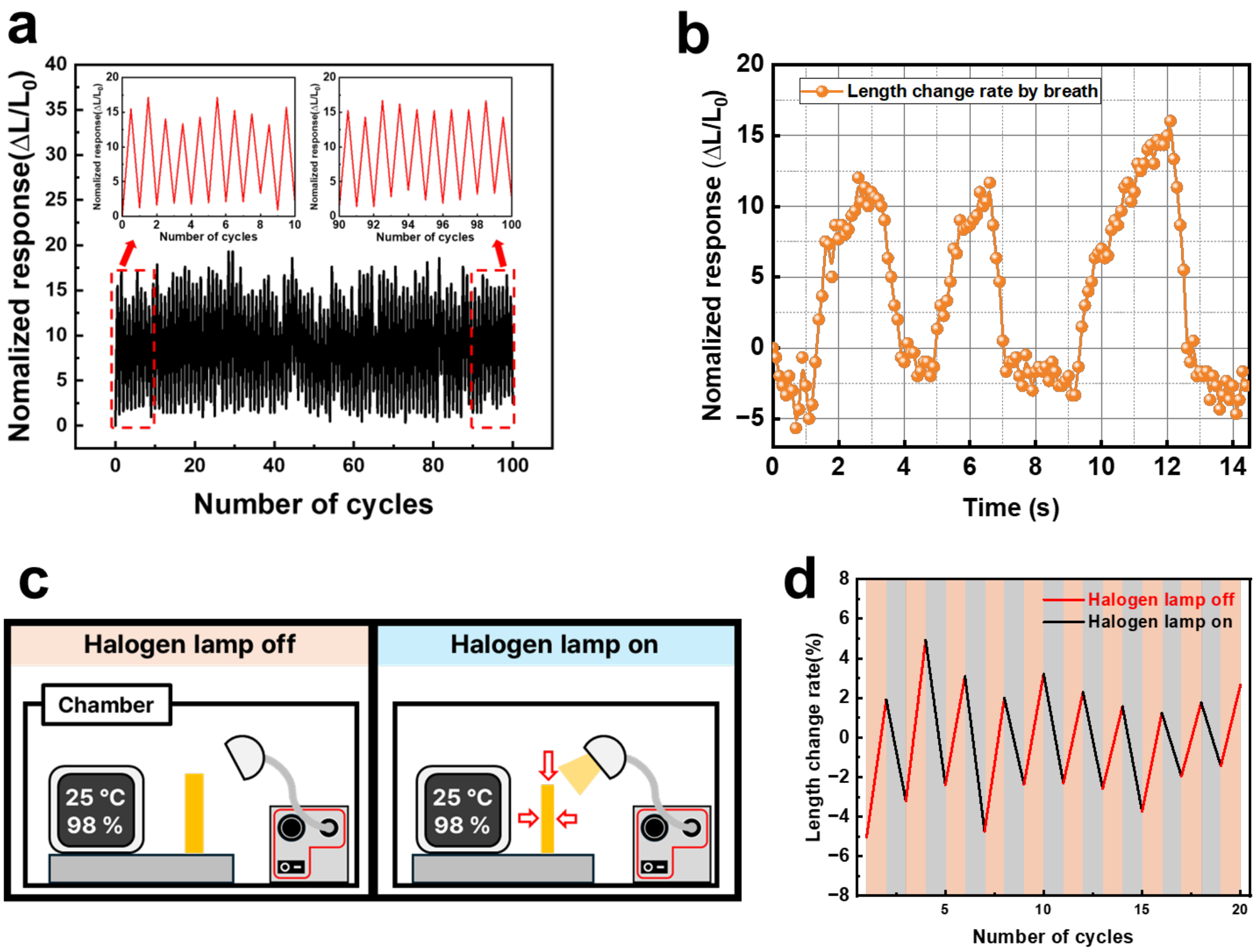

**Figure 4.** (a) Time-dependent length change of a bioinspired microactuator during repeated humidity cycling, showing reproducible expansion and contraction. (b) Magnified plot of the length change rate under repeated humidity variations induced by human breath. (c) Schematic illustration of the experimental setup for photothermal actuation under constant humidity (25 °C, RH 98%), comparing the halogen lamp off and on states. (d) Time-dependent length change of the microactuator during repeated photothermal on/off cycles, demonstrating reversible and repeatable actuation behavior.

The time trace in Figure 4b reveals that each expansion–contraction cycle occurs on the order of 2000 ms per cycle. In other words, the actuator can complete 0.5 expansion–contraction cycles per second. Such fast kinetics are notable for a polymer-based humidity actuator and likely arise from the microscale geometry and thin walls of the pillar, which accelerate moisture transport. The sharp, triangular pulses in the figure (highlighted by alternating shaded regions) show that the pillar length responds nearly instantaneously when RH is switched. In practical terms, the device exhibits sub-2000 ms response time, enabling dynamic actuation at 0.5 Hz. Figure 4c illustrates the experimental scheme for photothermal actuation under high humidity (RH 98%). The actuator is kept at a fixed 25 °C, 98% RH environment, and a halogen lamp is used to heat the pillar on demand. The lamp's broad-spectrum light is absorbed by the AuNP scaffold, which acts as an efficient photothermal transducer.[38-39] When the lamp is turned on, the gold network rapidly converts light to heat, locally raising the pillar temperature without changing ambient RH. The experimental data confirms that the actuator responds reversibly and reproducibly to the lamp stimulus (Fig. 4d). In each cycle, turning the lamp on (black trace) causes a sudden contraction of the pillar, and turning it off (red trace) leads to a return (expansion) to the original length. Under high humidity, the PVP phase absorbs water and remains in a swollen state. Upon photothermal heating, the local temperature rise reduces the equilibrium moisture content of PVP and accelerates water desorption. This thermally induced decrease in swelling leads to reversible contraction of the composite pillar, while the AuNP framework maintains structural integrity. Over ten consecutive on/off cycles, the black and red curves trace out nearly identical high and low states. There is no perceptible drift: the minima and maxima in each cycle recur at the same values. This shows that the contraction induced by heating is fully reversible, and the actuator recovers its swelling exactly when cooled. Compared to humidity-driven actuation, photothermal actuation enables rapid, localized, and on-demand control of deformation without globally altering the humidity, providing greater flexibility for device-level integration.

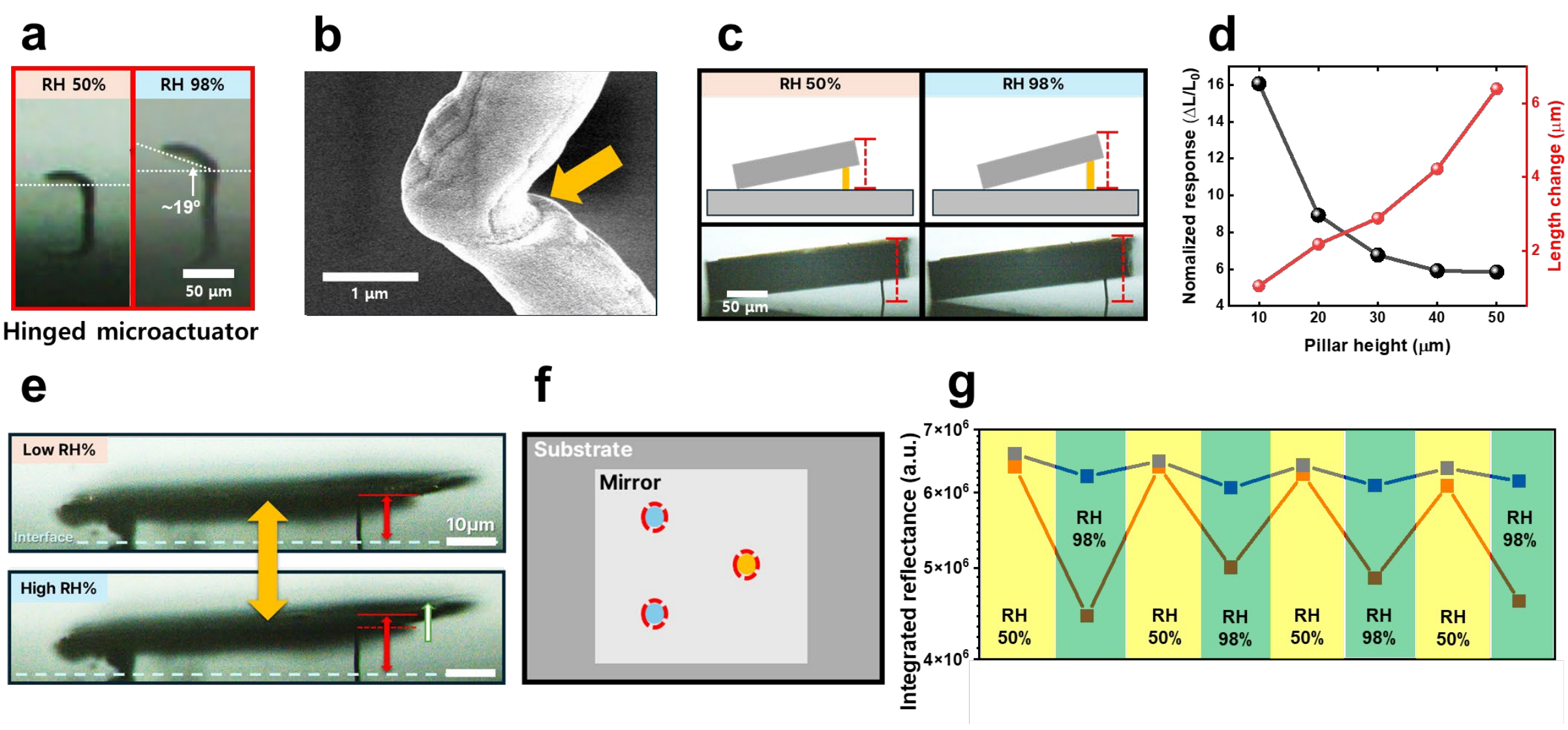

**Figure 5.** (a) Optical microscope images of a folded microactuator under low (RH 50%) and high (RH 98%) humidity, showing a humidity-induced change in the folding angle. (b) High-magnification SEM image of the hinge region in a folded microactuator, revealing a locally thinned geometry and a distinct contact line. (c) Schematic illustrations and corresponding optical microscope images of a loaded-lift configuration, where a microactuator lifts a rigid substrate under low and high humidity conditions. (d) Quantitative plots of absolute length

change and normalized response as a function of pillar height for the loaded-lift configuration. (e) Optical microscope images of a bioinspired microactuator lifting a metallic film under low and high humidity, showing humidity-dependent vertical displacement. (f) Top-view schematic illustration of the optical lever configuration supported by three pillars, including two thermally deactivated pillars and one active bioinspired microactuator. (g) Integrated visible reflectance (400–800 nm) from regions supported by thermally treated pillars (blue dot-line) and by bioinspired microactuators (red dot-line) under alternating humidity conditions (RH 50% and RH 98%).

In meniscus-guided 3D nanoprinting, it is possible to fabricate three-dimensional freestanding architectures with various geometries by controlling the motion direction of the motorized stage. Figure 5a shows an optical microscope image of a hinged microactuator with a folding angle of 90°. Unlike conventional straight microactuators, the hinged microactuator exhibits a humidity-dependent change in folding angle in response to the ambient relative humidity. Under high humidity, the AuNP–PVP composite tends to expand/elongate along its local axial direction due to PVP-driven moisture uptake. In a hinged geometry, however, the base is anchored to the substrate (approximately clamped or partially clamped), so the structure cannot fully relax the swelling-induced deformation through uniform axial extension. Instead, the imposed axial strain is mechanically converted into a bending moment at the hinge region, making rotation at the hinge an energetically favorable deformation mode. SEM imaging of the hinged microactuator reveals that the hinge region is noticeably thinner than the main body and exhibits a distinct contact line, implying a localized interfacial boundary rather than a fully homogeneous, continuous joint (Fig. 5b). This geometric and microstructural discontinuity effectively creates a compliant hinge with reduced bending stiffness compared to the thicker segments, thereby concentrating deformation at the hinge under humidity stimulation. This strain mismatch drives differential elongation across the folded legs and concentrates deformation at the kink, resulting in a measurable change in folding angle (Supplementary Movie S2).

The schematics and corresponding optical microscope images illustrate the loaded-lift actuation mode, in which the bio-inspired microactuator lifts a silicon wafer upon exposure to high humidity (Fig. 5c). As the relative humidity increases from 50% to 98%, the actuator undergoes axial swelling, resulting in a measurable vertical displacement of the wafer despite the applied load. In the loaded-lift configuration, both the absolute length change ($\Delta L$) and the normalized response ($\Delta L/L_0$) are significantly reduced compared to the free-standing case, indicating that the applied load suppresses the intrinsic swelling deformation of the actuator (Fig. 5d). When the pillar is in contact with a silicon wafer, a sustained compressive contact stress is imposed on the structure, which is known to lower the equilibrium swelling ratio and axial strain of hygroscopic polymer systems. As a result, the humidity-induced elongation of the pillar is partially inhibited, leading to a smaller $\Delta L$ and a substantially reduced normalized response. Notably, the normalized response in the loaded-lift mode saturates at a relatively short pillar height (~40 μm), whereas no clear saturation is observed up to 80 μm in the free-standing configuration. This early saturation suggests that, beyond a certain height, additional pillar length does not translate into increased effective actuation because the deformation becomes governed by a load-limited swelling equilibrium rather than by the available free swelling volume. In addition, for taller pillars under load, a growing fraction of the swelling-induced energy is likely redistributed into non-axial deformation modes, such as bending, tilting, or local compliance at the contact interface, rather than contributing to axial elongation. In other words, the normalized response reflects strain averaging over both actively swelling and constraint-dominated regions, rather than a simple scaling with the amount of hygroscopic material. Consequently, while free-standing actuators remain primarily limited by internal composite architecture, the loaded-lift actuation is dominated by external mechanical constraints, resulting in a reduced and rapidly saturating normalized response.

The mass of the pillar was estimated based on the molecular weight of PVP, the mass of individual gold nanoparticles, and the particle packing density observed in the SEM image (Fig. 2e). Using this approach, the mass of a microactuator with a height of 50 μm was estimated to be approximately 55 ng. The mass of the InP substrate used for the lifting experiment was approximately 100 μg. When torque effects are taken into account, the proposed bioinspired microactuator is therefore capable of lifting a load approximately 800 times its own weight. Figure 5e presents an optical microscope image showing a bioinspired microactuator lifting a metallic film, while Figure 5f illustrates a top-view schematic in which the metallic film is separated from the substrate by three supporting pillars. Among these, the two pillars marked in blue serve as mechanically rigid supports, in which actuation is suppressed by removing PVP through thermal treatment, whereas only the pillar marked in orange corresponds to the bioinspired microactuator that actively responds to humidity. This configuration enables unambiguous attribution of the observed optical changes to the actuation of the bioinspired microactuator. When the fabricated optical device is exposed to a high-humidity environment (RH = 98%), swelling of the microactuator induces a change in the height and tilt of the metallic film. (Supplementary Movie S3) Figure 5g quantifies the humidity-induced optical response by comparing the integrated visible reflectance from regions supported by the sintered AuNP pillars and by bioinspired microactuators. For the sintered AuNP pillars, which lack a hygroscopic component, the reflectance exhibits only a modest variation between RH 50% and RH 98%, remaining within approximately 5% on average. In contrast, regions supported by bioinspired microactuators show a pronounced and reversible modulation of reflectance, with an average change of approximately 33% upon humidity cycling. This large optical modulation originates from humidity-driven swelling of the bioinspired microactuator, which induces a measurable change in the height and tilt of the metallic film and consequently alters the reflection geometry. Overall, these results demonstrate that the bioinspired microactuator can efficiently convert a humidity stimulus into a substantial and repeatable optical signal, highlighting its potential for optical lever–based sensing and modulation applications.

## 3. Conclusion

In summary, this work demonstrates a novel microscale actuator design that merges a bioinspired graded composite architecture with a one-step meniscus-guided 3D printing process. By directly printing integrated gold–polymer pillars that mimic an insect joint's hard–soft interface, a monolithic actuator is produced that combines structural robustness and moisture-driven swelling without multi-step assembly. Systematic experiments – including comparative control tests and detailed microscopic characterization – provided mechanistic insight: PVP-only pillars collapsed under high humidity due to excessive softening, whereas Au-only pillars showed negligible deformation. In contrast, the composite pillars exhibited stable and reversible axial elongation, demonstrating that the nanoparticle scaffold provides mechanical reinforcement while the polymeric matrix supplies hygroscopic swelling. This study underscores the effectiveness of the bioinspired gradient design in achieving both strength and functionality at the microscale.

The combination of facile fabrication, robust performance, and insightful design opens broad opportunities for application. These hybrid microactuators could be integrated into future smart systems, from autonomous microrobots and environmental sensors to adaptive optical, wearable, and biomedical electronics. Their compatibility with additive manufacturing and microsystems integration makes them attractive for on-chip incorporation in microelectromechanical devices, flexible sensors, and lab-on-chip platforms. By bridging biomimetic mechanics and scalable manufacturing, this work paves the way toward responsive microdevices in next-generation electronic, optical, and hybrid systems..

## 4. Experimental Section/Methods

*Preparation of Ink*: The ink used for printing consisted of 10 nm Au nanospheres (polyvinylpyrrolidone (PVP)-coated; nanoComposix Inc.) dispersed in deionized water. The as-received colloidal solution had a gold concentration of 1 mg mL-1. Excess PVP was removed and the colloid was concentrated via centrifugation, yielding a final Au nanoparticle ink with a solid content of ~5 mg mL-1.

*Meniscus-guided 3D nanoprinting:* The printing system comprised three main components: (i) a manual micrometer stage (XYZ translation stage, Thorlabs), (ii) an automated three-axis stage (NanoMax equipped with stepper-motor actuators, Thorlabs), and (iii) a 10X objective lens (Mitutoyo) for real-time visual observation. Glass micropipettes were fabricated using a pipette puller (P-1000, Sutter Instrument), and the inner diameter of the pipette tip was adjusted to approximately 5 μm or 10 μm. For printing, the silicon substrate was mounted on the motorized XYZ stage, and the Au colloidal ink was loaded into the glass micropipette by capillary action. The ink-filled micropipette was fixed in a pipette holder positioned above the substrate. Using the manual stage, the pipette tip was first aligned and brought into focus under an optical microscope (camera). The substrate was then raised with the motorized stage until it contacted the pipette tip. Finally, after establishing contact between the pipette tip and the substrate, the motorized stage was translated downward along the vertical (z) axis at a controlled speed, enabling meniscus-guided 3D printing of nanoparticle architectures.

*Characterization*: After printing, the microstructure of the fabricated architectures was examined by field-emission scanning electron microscopy (FE-SEM). SEM imaging and energy-dispersive X-ray spectroscopy (EDS) were performed using a FE-SEM system (JSM-7900F, JEOL). The height variation of the structures was quantified from side-view optical microscopy images using ImageJ software. Reflectance changes of the metallic film deposited on top of the printed structures were measured using a Vis–NIR spectrometer (S4 7WB–UK, Ossila) coupled to an optical microscope.

*Finite-element simulation:* Finite-element simulations (COMSOL Multiphysics 6.1, Structural Mechanics module) were performed to compare the humidity-induced (RH50% to RH98%) deformation behavior of three pillar configurations with identical height. For modeling simplicity and computational efficiency, the pillars were approximated as two-dimensional structures, while retaining the essential mechanical constraints relevant to the experimental system. Three material configurations were considered: Au nanoparticle–PVP composite, PVP-only, and Au nanoparticle–only pillars. In all cases, the pillar height was fixed at 180 nm. For the Au NP–PVP composite model, Au nanoparticles were represented as circular structure with a diameter of 15 nm, embedded in a PVP matrix with an interparticle gap spacing of 3 nm, reproducing the experimentally observed nanoparticle packing. The PVP-only pillar was modeled as a homogeneous polymer film with identical height, while the Au NP-only pillar was constructed by closely stacking Au nanoparticles without PVP. A fixed constraint condition was applied to the base of the pillar to represent its adhesion to the substrate. The periodic boundary conditions were applied on right and left boundaries. Humidity-induced swelling was applied only to the PVP, while Au was treated as a non-swelling elastic material. The height displacements and first principal strains were analyzed.

**Acknowledgements**

S-J.E, V.D. and S.K contributed equally to this work. This research was supported by the Basic Science Research Program through the National Research Foundation of Korea (NRF) (grant no. RS-2023-00219703). This research was supported by NRF Sejong Science fellowship (NRF-2021R1C1C2011447). V.D and T.Z acknowledge the support from the German Ministry of Education and Research (BMBF) within the PhoQuant project (grant number 13N16103).

**Conflict of Interests**

The authors declare no conflict of interests.

**Data Availability Statement**

The data supporting the findings of this work are available from corresponding authors upon reasonable request.

**Supporting Information**

Supporting Information (S1 – S3 movies).